\documentstyle[11pt,aaspp4]{article}

\begin{document}

\title{The Expected Rate of Gamma-Ray Burst Afterglows In Supernova
Searches}
\author{Eric Woods$^1$ and Abraham Loeb$^2$}
\medskip
\affil{Astronomy Department, Harvard University, 60 Garden St.,
Cambridge, MA 02138}
\altaffiltext{1}{email: ewoods@cfa.harvard.edu}
\altaffiltext{2}{email: aloeb@cfa.harvard.edu}

\begin{abstract}

We predict the rate at which Gamma-Ray Burst (GRB) afterglows should be
detected in supernova searches as a function of limiting flux.  Although
GRB afterglows are rarer than supernovae, they are detectable at greater
distances because of their higher intrinsic luminosity.  Assuming that GRBs
trace the cosmic star formation history and that every GRB gives rise to a
bright afterglow, we find that the average detection rate of supernovae and
afterglows should be comparable at limiting magnitudes brighter than
$K=18$.  The actual rate of afterglows is expected to be somewhat lower
since only a fraction of all $\gamma$--ray selected GRBs were observed to
have associated afterglows.  However, the rate could also be higher if the
initial $\gamma$--ray emission from GRB sources is more beamed than their
late afterglow emission. Hence, current and future supernova searches can
place strong constraints on the afterglow appearance fraction and the
initial beaming angle of GRB sources.

\end{abstract}

\keywords{gamma rays: bursts}

\section{Introduction}

Since their discovery in the late 1960's (Klebasadel et al. 1973) through
early 1997, Gamma-Ray Bursts (GRBs) had defied all attempts to determine
their distance scale conclusively.  The Burst And Transient Source
Experiment (BATSE) on board the Compton Gamma-Ray Observatory (GRO) showed
that the burst population is highly isotropic (Meegan et al. 1993; Briggs
et al. 1993), suggesting that bursts occur at cosmological distances or in
an extended Galactic halo. Moreover, the cumulative number counts of faint
bursts deviated from that of a uniform distribution of sources in Euclidean
space and flattened at faint fluxes, consistent with the expected effect
of a cosmological redshift (Fishman \& Meegan 1995, and references
therein).  Last year, with the advent of the BeppoSAX satellite (Boella et
al.  1997), it became possible to localize GRB sources to within an
arcminute on a timescale of hours.  Such fast, accurate localizations were
quickly followed by the detection of delayed X-ray (Costa et al.  1997),
optical (van Paradijs et al. 1997), and radio (Frail et al.  1997)
counterparts to GRB sources.  In particular, FeII and MgII absorption lines
were detected at a redshift $z=0.835$ in the spectrum of the optical
counterpart to GRB970508 (Metzger et al. 1997), demonstrating conclusively
that this burst occurred at a cosmological distance with a redshift
$z>0.835$.  The isotropy of the burst population and the flattening of
their number counts, taken in combination with the fact that the first
confirmed redshift for an optical counterpart is high, provides strong
evidence that GRB sources are located at cosmological distances.

Most plausible GRB models involve either the collapse of a single massive
star (e.g. Usov 1992; Woosley 1993; Paczy\'nski 1998), or the coalescence
of two compact objects -- two neutron stars or a neutron star and a black
hole -- in a binary system (e.g.  Paczy\'nsky 1986; Eichler et al. 1989;
Narayan et al.  1992; Mochkovitch et al. 1993; Rees 1997).  Since the
lifetime of these progenitors is short compared to the Hubble time at a
redshift $z\la 5$, the cosmic GRB rate should simply be proportional to the
star formation rate at these redshifts, without any appreciable delay due
to the finite progenitor lifetime.  The cosmic rate of massive star
formation rate has been determined from the $U$ and $B$-band luminosity
density in Hubble Deep Field (Madau et al.  1996; Madau 1996; Madau,
Pozzetti, \& Dickinson 1997; Madau 1997).  The inferred star formation rate
$\dot\rho_{\rm s}(z)$ can then be converted to a GRB explosion rate $R_{\rm
GRB}(z)$, based on the requirement that the latter would fit the observed
number count distribution of $\gamma$--ray selected GRBs (Wijers et al.
1997).

Cosmological GRBs are at least $10^4$ times rarer than Type II supernovae
(SNeII) -- possibly even $\sim 10^6$ times rarer if GRBs occur primarily at
high redshifts following the cosmic star formation history (Wijers et al.
1997).  However, at peak luminosity, the GRB afterglows are $\sim
10^3$--$10^4$ times brighter than SNeII.  In Euclidean space, this would
imply that GRBs are detected from a volume bigger by a factor $\sim
(10^{4})^{3/2}= 10^6$, roughly canceling out the factor by which they are
rarer than supernovae.  Hence we expect that at some relatively bright
limiting flux, the rate of afterglow detections should become comparable to
that of SN detections.  Current and future supernova searches should
provide information about the fraction of GRBs which produce detectable
afterglows. The statistics of bursts in 1997 for which afterglows could
have been identified implies that this fraction is of order tens of percent
(e.g., Castro-Tirado 1998).  On the other hand, there could also be a
population of afterglows without a GRB precursor.  This would occur if the
source emits a jet from which the $\gamma$--ray emission is more beamed
than the subsequent optical afterglow radiation due to the deceleration of
the jet by the ambient gas and the corresponding decline in its
relativistic beaming with time (Rhoads 1997).  A jet geometry would imply a
higher rate of afterglow detections in supernova searches.

In this {\it Letter}, we predict the detection frequency of GRB afterglows
as a function of limiting flux at various observed wavelengths, and compare
this rate with the analogous predictions for SNe Type Ia and Type II at
high redshifts.  We assume throughout a flat, $\Omega =1$, $\Lambda =0$,
cosmology, with a Hubble constant $H_0=50$ km s$^{-1}$ Mpc$^{-1}$.

\section{Input Parameters}

The inferred energy release in cosmological GRBs and their afterglows is
comparable to the binding energy of a neutron star, $\sim 10^{53}$ ergs.
We therefore assume that GRBs, like supernova explosions, have evolved
stars as their progenitors.  Thus, the cosmic GRB and SN rates should both
trace the cosmic star formation history.  For progenitors with an effective
lifetime $\tau$, the corresponding event rate at time $t>\tau$ will be
proportional to the star formation rate at time $(t-\tau)$.  Following Dwek
(1997), we get for the rate of supernovae Type Ia (SNeIa) per comoving
volume,
\begin{equation}
R_{\rm SNIa}(t) = {{\beta \int_{3}^{16} \phi(m_{\rm b})dm_{\rm b}
\int_{\mu_{\rm m}}^{0.5} \dot\rho_{\rm s}[t-\tau(\mu m_{\rm b})] 
f(\mu) d\mu} \over{\int_{m_{\rm l}}^{m_{\rm u}} m\phi(m)dm}},
\label{SNIRate}
\end{equation}
where $\dot\rho_{\rm s}$ is the stellar mass formed per unit comoving
volume per unit time, $m$ denotes masses in solar units, $\phi(m_{\rm b})$
is the initial mass function (IMF) for binary systems with total mass in
the range $m_{\rm l}<m_{\rm b}<m_{\rm u}$, $f(\mu)=24\mu^2$ is the
distribution of secondary-to-total-binary-mass ratios $\mu\equiv m/m_{\rm
b}$, $\tau(m)$ is the lifetime of a progenitor star of mass $m$, and
$\beta<1$ is the SNIa rate amplitude, an adjustable parameter.  We find
that $\beta=0.05$ yields the best fit to the local and high-redshift
supernova observations, and adopt this value.  The lower limit in the
second integral of the numerator is $\mu_{\rm m}={\rm max}[m(t)/m_{\rm
b},(m_{\rm b}-8)/m_{\rm b}]$, where $m(t)$ is the mass of stars which are
turning off the main sequence at time $t$, i.e. the inverse of $\tau(m)$
(Greggio \& Renzini 1983).  This lower limit take into account the fact
that stars more massive than the turnoff mass should not be included in the
integral.  We assume a Salpeter (1955) IMF, $\phi(m)\propto m^{-2.35}$,
with $m_{\rm l}=0.1$ and $m_{\rm u}=125$.  For Type II supernovae, the
progenitor lifetime is negligible compared to the Hubble time, and so we
take $\tau=0$,
\begin{equation}
R_{\rm SNII}(t) = {{\dot\rho_{\rm s}(t) [\int_{16}^{m_{\rm u}} \phi(m) dm 
+ (1-\beta)\int_8^{16}\phi(m)dm]}\over{\int_{m_{\rm l}}^{m_{\rm u}} m\phi(m)dm}}.
\label{SNIIRate}
\end{equation}
The second term in the numerator accounts for single stars and members of
binary systems which do not lead to Type Ia supernovae.  

If GRBs have stellar progenitors, then their rate also traces the star
formation rate.  This scenario was investigated by Wijers et al.  (1997),
who also assumed that the progenitors are short-lived compared to the
Hubble time ($\tau=0$), and derived a best-fit constant of proportionality
between the GRB rate $R_{\rm GRB}$ and the star formation rate
$\dot\rho_{\rm s}$.  Given our assumptions about the IMF, their result
translates to
\begin{equation}
R_{\rm GRB}(t) = 7.1\times 10^{-7} R_{\rm SNII}(t).
\label{GRBRate}
\end{equation}
Note that this rate is substantially lower (by a factor $\sim 150$) than
the best-fit rate derived by assuming a non-evolving burst population,
$R_{\rm GRB}\sim 10^{-4}R_{\rm SNII}$.  However, Wijers et al. (1997) also
obtain a best-fit GRB standard-candle luminosity of $1.6\times 10^{52}$ erg
s$^{-1}$ (assuming $\Omega =1$, $H_0=50~{\rm km~s^{-1}~Mpc^{-1}}$), which
is $\sim 20$ times brighter than that obtained for a non-evolving source
population.  In the following we will consider the best-fit rates for both
GRB sources which are evolving (following the cosmic star formation
history) or non-evolving.

Next, let us consider a population of transient sources, which are standard
candles in peak flux and are characterized by a comoving rate per unit
volume $R(z)$.  The observed number of new events per unit time brighter
than flux $F_\nu$ at observed wavelength $\lambda$ for such a population is
given by
\begin{equation}
\dot{N}(F_\nu;\lambda) = \int_0^{z_{{\rm max}(F_\nu;\lambda)}} R(z) (1+z)^{-1}
(dV_{\rm c}/dz) dz,
\label{CountsPerYear}
\end{equation}
where $z_{\rm max}(F_\nu,\lambda)$ is the maximum redshift at which a
source will appear brighter than $F_\nu$ at an observed wavelength
$\lambda=c/\nu$, and $dV_{\rm c}$ is the cosmology-dependent comoving
volume element. The above integrand includes the $(1+z)$ reduction in the
apparent rate due to the cosmic time dilation.  For $\Omega=1$,
$\Lambda=0$, the comoving volume element covered by solid angle $d\Omega$
and redshift interval $dz$ is given by
\begin{equation}
dV_{\rm c} = 4\left({c\over{H_0}}\right)^3 
{{(1+z-\sqrt{1+z})^2}\over{(1+z)^{7/2}}}~d\Omega dz .
\label{Dvdz}
\end{equation}

Equation~(\ref{CountsPerYear}) is appropriate for a threshold experiment,
such as BATSE, which monitors the sky continuously and triggers when the
detected flux exceeds a certain value, and hence identifies the most
distant sources only when they are near their peak flux.  For search
strategies which involve taking a series of ``snapshots'' of a field and
looking for variations in the flux of sources in successive images, one
does not necessarily detect most sources near their peak flux.  In this
case, the {\it total} number of events ({\it not} per unit time) brighter
than $F_\nu$ at observed wavelength $\lambda$ is given by
\begin{equation}
N(F_\nu;\lambda) = \int_0^\infty R(z) t_\star (z;F_\nu,\lambda) (dV_{\rm
c}/dz) dz,
\label{Counts}
\end{equation}
where $t_\star (z;F_\nu,\lambda)$ is the rest-frame duration over which an
event will be brighter than the limiting flux $F_\nu$ at redshift $z$.
This is a naive estimate of the so-called ``control time''; in practice,
the effective duration over which an event can be observed is shorter,
owing to the image subtraction technique, host galaxy magnitudes, and a
number of other effects which reduce the detection efficiency (Pain et al.
1996).

In applying equations~(\ref{SNIRate})--(\ref{Counts}), we use the star
formation rate $\dot\rho_{\rm s}(z)$ determined by Madau (1997) and convert
cosmic time to redshift according to the standard cosmological formulae.
For the ages of the SNIa progenitors we use the calculations by Schaller et
al. (1992).  To determine the maximum redshift $z_{\rm max}$ and the
effective duration $t_\star$ for both SNe and GRBs, we use the
solar-metallicity, time-dependent, spectral models of Eastman et al. (1994)
for SNeII, the template lightcurves of Riess, Press, \& Kirshner (1996)
together with the spectral models of Nugent et al. (1997) for SNeIa --
extended out to $2.2~\mu$m with a simple blackbody fit, and the broken
power law model of Waxman (1997) for GRB afterglows.  In all cases, we
truncate the emission spectrum at rest wavelengths shorter than $\sim 10^3$
\AA, to reflect the absorption by HI  beyond the Lyman-limit in the 
local environment of the source and the intergalactic medium.  For the
afterglows, we modify Waxman's choice of parameters after peak luminosity
so as to match the faster decay slope observed for GRB970228 (Galama et al.
1997; Fruchter et al. 1998) and GRB970508 (Galama et al. 1998), namely
$L_\nu\propto t^{-1.1}$. We also introduce a cutoff time $t_{\rm cut}$
beyond which we truncate the afterglow flux so that the source counts will
converge (see \S 3 for the effect of varying $t_{\rm cut}$).  The resulting
GRB spectrum as a function of time is
\begin{equation}
L_\nu(t) = 2.94\times 10^{30} \eta \left({{t_{\rm
day}}\over{t_\lambda}}\right)^\alpha {\rm erg{\ }s}^{-1} {\rm Hz}^{-1},
\label{GRBSpec}
\end{equation}
where $t_{\rm day}\equiv (t/{\rm day})$, $\alpha=0.5$ for $t_{\rm day}\leq
t_\lambda$, $\alpha=-1.1$ for $t_\lambda<t_{\rm day}<t_{\rm cut}$, and
$t_\lambda\equiv (1.41\times 10^{14}{\rm Hz}/\nu)^{2/3}~{\rm days} =
(\lambda/2.13\mu{\rm m})^{2/3}$ days.  The parameter $\eta$ reflects the
choice of a typical GRB redshift, with $\eta =1$ for a non-evolving burst
population -- corresponding to a typical redshift $z\sim 1$.  On the other
hand, Wijers et al. (1997) derive a characteristic $\gamma$--ray luminosity
which is $\sim 20$ times higher than that obtained for a non-evolving
population. The afterglow luminosity depends on the total fireball energy
(Waxman 1997), which is proportional to the product of the $\gamma$--ray
luminosity and the rest-frame duration.  Taking all of these dependences
into account, we derive $\eta\sim 10$ for a population which traces the
star formation history.  We express the afterglow emission in terms of the
luminosity per unit frequency, since detector sensitivities are often
expressed in Janskys (1 Jy = $10^{-23}$ erg s$^{-1}$ cm$^{-2}$ Hz$^{-1}$).
From equation~(\ref{GRBSpec}), we have for GRB afterglows
\begin{eqnarray}
t_\star (z;F_\nu,\lambda) &=& {\rm min}\{t_{\rm cut},
A(z;F_\nu)^{-0.9}t_{\lambda/(1+z)}\}-
A(z;F_\nu)^2 t_{\lambda/(1+z)},\nonumber\\
A(z;F_\nu)&\equiv& {{4.0\times 10^2\eta^{-1}}\over{(1+z)}}
\left[{F_\nu\over{{\rm Jy}}}\right]
\left[{{D_{\rm L}(z)}\over{{\rm Gpc}}}\right]^2\nonumber\\
\label{GRBtstar}
\end{eqnarray}
where $D_{\rm L}(z)$ is the cosmology-dependent luminosity distance,
$D_{\rm L}(z)=(2c/H_0)(1+z-\sqrt{1+z})$ for $\Omega=1$ and $\Lambda=0$.

\section {Results}

We apply equation~(\ref{CountsPerYear}) to determine the average rate at
which GRB afterglows, SNeIa, and SNeII should be detected in a field of 1
deg$^2$ as a function of limiting flux.  Figure 1 shows the results for
different spectral bands, namely the $K$ ($\lambda_{\rm eff}=2.2$ $\mu$m),
$R$ ($\lambda_{\rm eff}=7000$ \AA), $B$ (effective wavelength,
$\lambda_{\rm eff}=4400$ \AA), and $U$ ($\lambda_{\rm eff}=3650$ \AA)
bands.  The $B$-band rate of $\sim 4800$ SNeII yr$^{-1}$ deg$^{-2}$
brighter than 1 nJy is slightly lower than a recent prediction (Madau 1998)
of 25--35 SNeII yr$^{-1}$ per $4'\times 4'$ field of view of the Next
Generation Space Telescope (NGST, Mather \& Stockman 1996), where the range
of values corresponds to a range of assumptions about the amount of dust
extinction between the observer and the sources.  Our slight underestimate
of $\sim 21$ SNeII yr$^{-1}$ per NGST field results from the fact that we
do not correct the star formation rate for the effects of dust extinction.
Our predictions from equations~(\ref{SNIRate})--(\ref{SNIIRate}) for the
local ($z=0$) SNIa and SNII rates are consistent with recent observations
(Cappellaro et al. 1997; Sadat et al. 1998).  The SNIa rate we obtain at a
redshift $z\sim 0.4$ agrees to within statistical uncertainties with the
rate determined from observations of high-redshift SNeIa (Pain et al.
1996).  The Type Ia supernova counts overtake the Type II counts at
relatively bright fluxes, since SNeIa are an order of magnitude brighter
than SNeII and hence dominate the counts in a shallow magnitude-limited
sample, even though their absolute rate is a fraction ($\sim 30$--$50\%$)
of the Type II rate.  For both types of supernovae, we truncate the
spectrum at rest wavelengths shorter than $\sim 10^3$ \AA~due to the
expected galactic and intergalactic absorption by neutral hydrogen.

The predicted GRB afterglow rate in Figure 1 is nearly independent of the observed
band, since the peak GRB flux in equation~(\ref{GRBSpec}) is the same at
all wavelengths.  The sharp ``knee'' at $10^{-4}$--$10^{-3}$ Jy reflects
the edge in the spatial distribution of luminous bursts due to the
non-Euclidean geometry of the Universe and the sharp decline in the star
formation rate at $z>4$.  We extrapolate the star formation rate using a
power-law dependence on $(1+z)$ at $z>4$.
Our extrapolation is consistent with the upper limit on $\dot\rho_{\rm
s}(z)$ around $z=5.5$, based on the lack of ``$V$ dropout'' galaxies in the
Hubble Deep Field (Madau 1996).  Production of bright afterglows, of the
type observed in GRB970228 or GRB970508, requires an ambient gas density of
$\sim 1~{\rm cm^{-3}}$ (Waxman 1997), which is typical of the interstellar
medium of disk galaxies.  Hence, we expect significant Lyman-limit
absorption due to HI in the local burst environment, and so we truncate the
afterglow spectrum at rest wavelengths $<10^3$ \AA.  This explains the
variation with observed wavelength of the rate of faint afterglows.
%At fluxes fainter
%than $10^{-4}$ Jy, we are probing emission beyond the Lyman limit from
%$z>4$ and so we are no longer picking up any more sources.
Note that at
all observed wavelengths, the afterglow rate becomes comparable to the Type
II SN rate at magnitudes $\sim 18-20$.  
This means that if afterglows occur with the same frequency as the
$\gamma$--ray selected GRBs, then they should turn up with
comparable numbers to supernovae in searches of this depth.

Figure 2 shows the results of applying equation~(\ref{Counts}) to obtain
the number of events that are detectable in a single ``snapshot'', taking
into account the effective length of time during which each type of event
should be visible at a given sensitivity.  Our prediction for the $R$-band
counts of SNeIa is consistent with the results of the high-redshift survey
of Pain et al. (1996), who found three Type Ia supernovae in the range
$21.3<R<22.3$ across a 1.73 deg$^2$ area on the sky.  Our predicted rate is
somewhat higher, due to the fact that our naive estimate of the ``control
time'' $t_\star$ assumes 100\% efficiency over the duration of the event,
while the actual efficiency (and the corresponding value of $t_\star$) are
lower in a real search. Afterglows are less abundant relative to supernovae
in Figure 2 as compared to Figure 1, because the timescale for the decline
of their lightcurve around peak flux is an order of magnitude shorter [cf.
Eqs.~(\ref{CountsPerYear}) and~(\ref{Counts})].

The lower and upper solid lines in Figures 1 and 2 show the predicted
afterglow rates with and without evolution.  The two lines differ
substantially at faint fluxes and coincide to within a factor of a few at
bright fluxes.  The coincidence at the bright end results from the fact
that the adopted GRB luminosities and rates were chosen so as to reproduce
the same flux distribution of BATSE--selected GRBs.  However, at faint
fluxes which require sensitivities greater than that achieved by the BATSE
experiment, the counts converge to different values. The no-evolution model
predicts many more bursts at redshifts low enough such that the observed
frequency falls below the Lyman limit at emission.

In order to apply equation~(\ref{Counts}) to the afterglow population, we
needed to assume that the afterglow emission truncates after some time
$t_{\rm cut}$, since otherwise the power-law lightcurve implied by
equation~(\ref{GRBSpec}) would lead to an untruncated power-law in the
number counts, and a divergence of the number of afterglows at faint
fluxes.  Observations indicate that GRB afterglows continue their power-law
decline, $L_\nu
\propto t^{-1.1}$, over a considerable fraction of a year (Galama et al.
1997), and so in Figure 2 we assumed a value of $t_{\rm cut}=1$ year.  The
effect of varying this choice of $t_{\rm cut}$ is illustrated in Figure 3;
the position of the number count ``knee'' shifts to fainter fluxes when
$t_{\rm cut}$ is increased since a larger $t_{\rm cut}$ corresponds to a
larger effective volume being surveyed.  The uncertainty concerning $t_{\rm
cut}$ translates to an uncertainty in the expected faint counts.  By
contrast, the afterglow rate predictions displayed in Figure 1 are
effectively independent of the detailed shape of the lightcurve.

\section {Conclusions}

We have predicted the rate at which GRB afterglows should be detected in
supernova searches as a function of limiting flux.  Requiring that the GRB
population would evolve according the cosmic star formation history reduces
the number counts of afterglows at faint fluxes by $\sim 2$ orders of
magnitude relative to the no-evolution case (see the solid lines in Figs. 1
and 2).  Our main result is that the average detection rate of new
afterglows (in both evolutionary cases) is comparable to that of supernovae
for a continuous search with a magnitude limit of $K=18$ (cf. Fig. 1),
assuming that every GRB produces a bright afterglow of the type observed
for GRB970228 or GRB970508.  However, the {\it number} of detectable
afterglows in a ``snapshot'' search (cf. Fig. 2) could be an order of
magnitude smaller than the number of detectable supernovae, since the
afterglow emission declines around its peak flux an order of magnitude
faster.  The actual number of afterglows in real surveys might be even
lower since only a fraction of all $\gamma$--ray selected GRBs detected in
1997 were observed to give rise to bright afterglows. (One theoretical
interpretation of this result could be in terms of the proportionality
between the predicted afterglow luminosity and the square root of the
ambient gas density (e.g., Waxman 1997).  The distribution of afterglow
luminosities is expected to reflect the range of gas densities in GRB
environments and might span several orders of magnitude.) The reduction
factor due to the small appearance probability of bright afterglows might
be counteracted by a possible enhancement factor due to beaming.  If the
initial gamma-ray emission is beamed to an angle $\sim1/\gamma_0$ where
$\gamma_0$ is the initial Lorentz factor, while the afterglow radiation is
produced when the fireball has been decelerated to a modest Lorentz factor
$\gamma_{\rm a}$, then the number of optically-selected afterglows could be
greater by a factor $\sim (\gamma_0/\gamma_{\rm a})^2$ than the number of
$\gamma$--ray selected GRBs (Rhoads 1997).  Since typical bursts have
$\gamma_0\ga 10^{2-3}$ (e.g., Fenimore, Epstein, \& Ho 1993; Woods
\& Loeb 1995) while the optical afterglow emission
occurs at $\gamma_{\rm a}\sim 10^{1-2}$, this could boost the expected
afterglow rates by up to four orders of magnitude.  Existing supernova
searches (e.g., Garnavich et al.  1998; Perlmutter et al.  1998) reach an
effective magnitude limit $R\ga 22$ (Kirshner 1998), and could already
place meaningful constraints on the beaming of the $\gamma$--ray emission.
Future searches might also constrain the appearance probability of the
afterglow emission and the evolution of the GRB population.

The Next Generation Space Telescope, with its detection sensitivity of
$\sim 10^{-9}$ Jy at 1--3.5$\mu$m (Mather \& Stockman 1996), will be able
to see GRB afterglows out to redshifts $z\ga 10$, if they exist\footnote{If
GRBs result from binary coalescence events of compact stars, then their
abundance might be suppressed as long as the age of the Universe is shorter
than the characteristic coalescence time. On the other hand, the initial
mass function of the first stars is likely to be tilted towards high mass
stars, which are probable progenitors of neutron stars or black holes.
This would enhance the abundance of such binaries per generation of
stars.}.  Afterglows might be the brightest sources at these redshifts
aside from early quasars (Haiman \& Loeb 1997b).  Based on the expected
supernova rate at $z\ga5$ (Miralda-Escud\'e \& Rees 1997), there might be
one detectable afterglow per year per $\sim {\rm deg}^2$ in the
no-evolution model.  The stretching of the observed lifetime of afterglows
from high redshifts and the potential enhancement factor for the afterglow
rate due to beaming, might bring the detection of afterglows closer to the
realm of feasibility.  The discovery of high redshift afterglows could then
be used to trace the star formation history out to the ``dark ages'' of the
Universe when the first stars and quasars formed (e.g., Rees 1996; Haiman
\& Loeb 1997a,b; Loeb 1997). The optical afterglow radiation could also be
used as a means of identifying the reionization redshift through the
spectral location of the Gunn-Peterson absorption trough which is produced
by the neutral intergalactic medium prior to reionization.

\acknowledgements

We thank Bob Kirshner for useful discussions and Ron Eastman for providing
us with the latest model spectra for Type II supernovae.

%\vfill\eject

%\centerline{\bf Figure Captions}

%\bigskip

%\noindent
%{\bf Fig. 1 --} Cumulative observed rate $\dot{N}(>F_\nu)$ per year per
%square degree of GRB afterglows ({\it solid curves}) at four wavelengths,
%corresponding to the $K$, $R$, $B$, and $U$ bands.  The lower solid curve
%assumes the best-fit rate and luminosity for GRB sources which trace the
%star formation history (Wijers et al. 1997), while the upper solid curve
%assumes the best-fit values for a non-evolving GRB population.  For
%comparison, we show the counts for SNe Ia ({\it dashed curves}) and
%SNe II ({\it dotted curves}).

%\noindent
%{\bf Fig. 2 --} Cumulative number counts $N(>F_\nu)$ per square degree for
%GRB afterglows, with a cutoff time of $t_{\rm cut}=1$ yr.  Notations are
%the same as in Figure 1.

%\noindent
%{\bf Fig. 3 --} Cumulative $B$-band counts of GRB afterglows for various
%values of the cutoff time $t_{\rm cut}$.  From top to bottom, the curves
%are for $t_{\rm cut}$ = $\infty$, 10 years, 1 year, and 1 month.
%\vfill\eject

\begin{figure}
\plotone{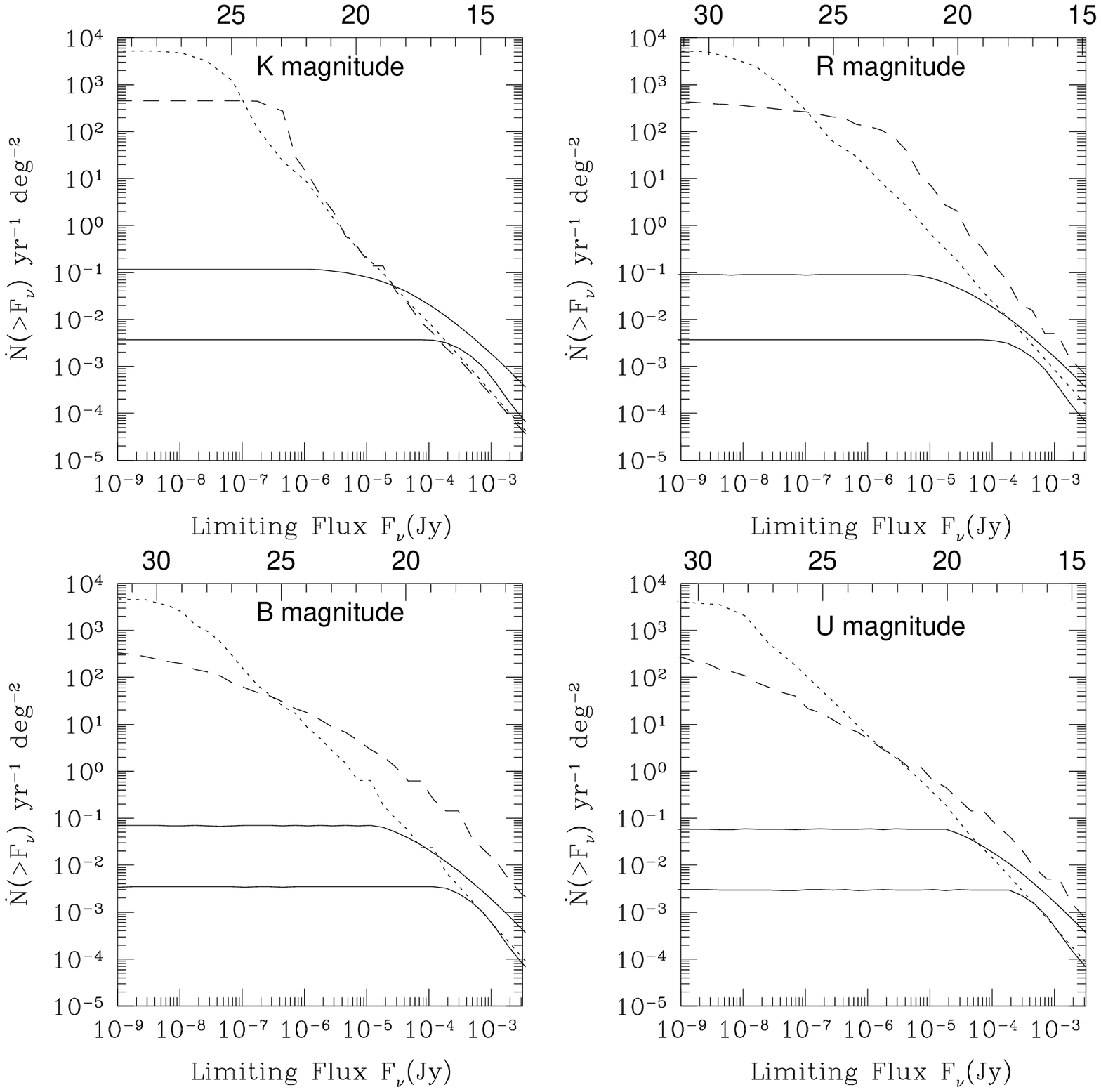}
\caption{Cumulative observed rate $\dot{N}(>F_\nu)$ per year per
square degree of GRB afterglows ({\it solid curves}) at four wavelengths,
corresponding to the $K$, $R$, $B$, and $U$ bands.  The lower solid curve
assumes the best-fit rate and luminosity for GRB sources which trace the
star formation history (Wijers et al. 1997), while the upper solid curve
assumes the best-fit values for a non-evolving GRB population.  For
comparison, we show the counts for SNe Ia ({\it dashed curves}) and
SNe II ({\it dotted curves}).
}
\end{figure}

\begin{figure}
\plotone{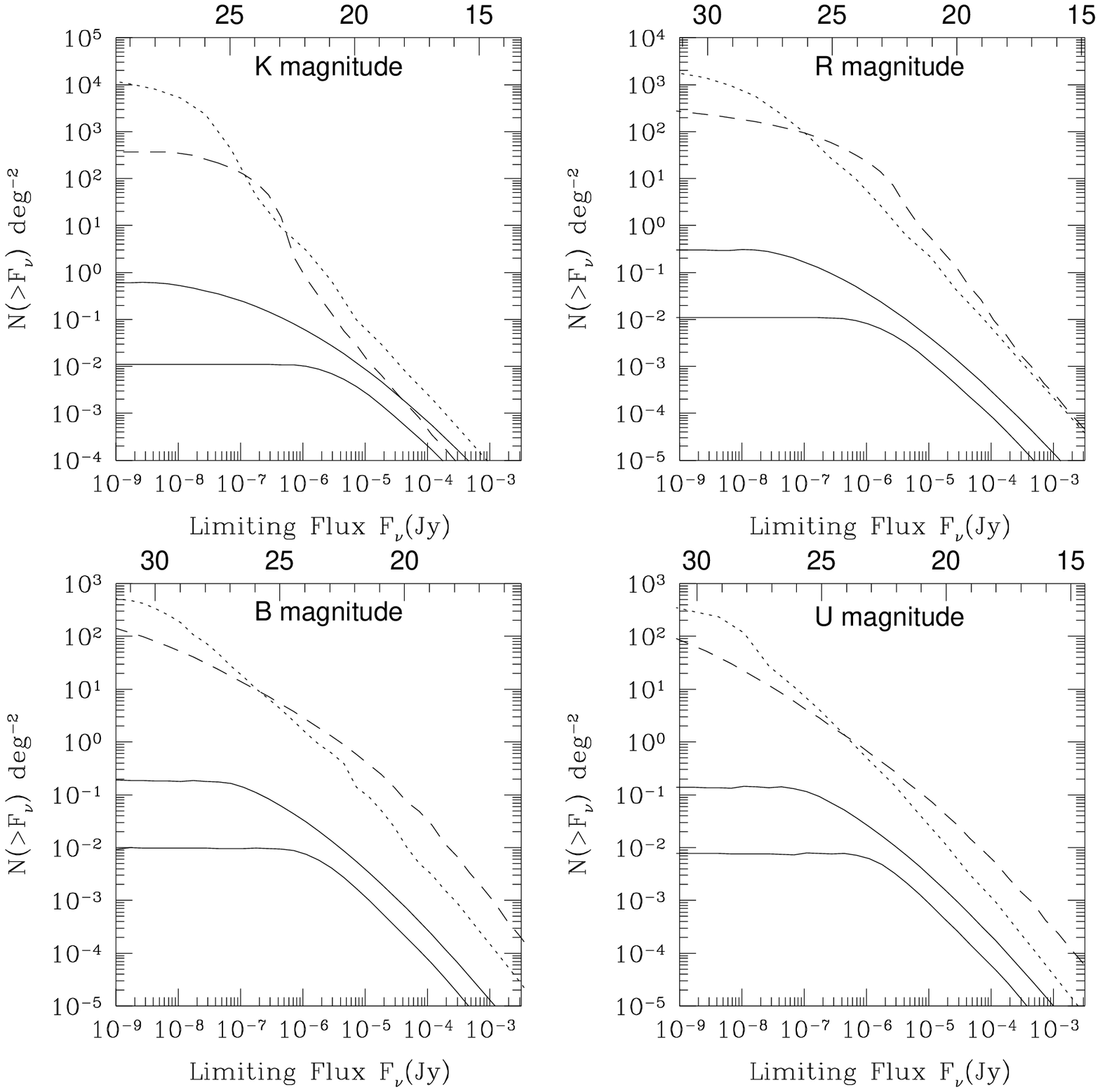}
\caption{Cumulative number counts $N(>F_\nu)$ per square degree for
GRB afterglows, with a cutoff time of $t_{\rm cut}=1$ yr.  Notations are
the same as in Figure 1.}
\end{figure}

\begin{figure}
\plotone{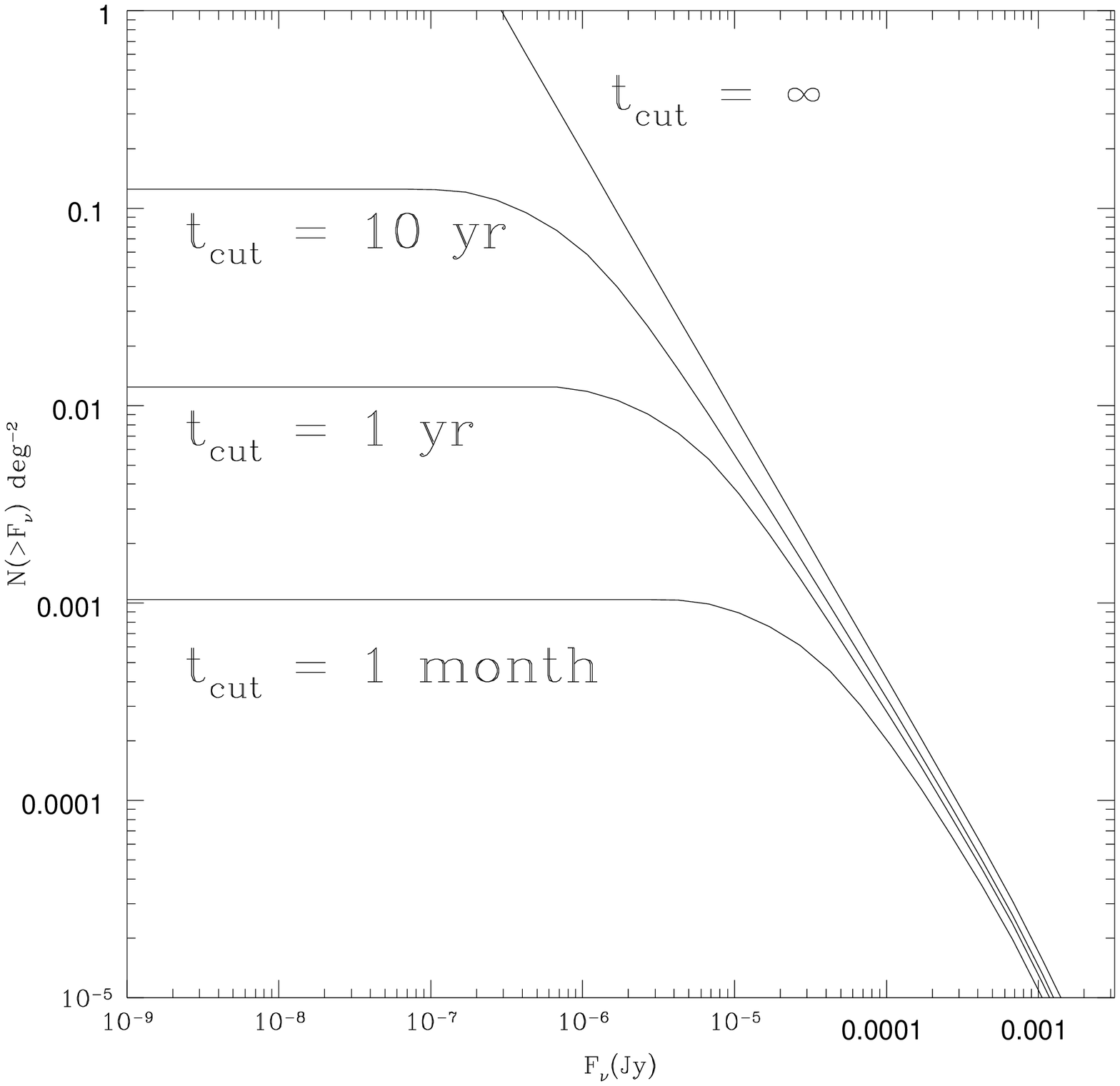}
\caption{Cumulative $B$-band counts of GRB afterglows for various
values of the cutoff time $t_{\rm cut}$.  From top to bottom, the curves
are for $t_{\rm cut}$ = $\infty$, 10 years, 1 year, and 1 month.}
\end{figure}


\begin{references}

\reference{} Boella, G., et al. 1997, A\&AS, 122, 299
\reference{} 
Briggs, M. S., et al. 1993, in Proc. of the Huntsville Gamma-Ray Burst
Workshop, ed. G. Fishman, J. Brainerd, \& K. Hurley (New York: AIP), 44
\reference{} Cappellaro, E., Turatto, M., Tsvetkov, D. Yu., Bartunov, O. S., 
Pollas, C., Evans, R., \& Hamuy, M. 1997, A\&A, 322, 431
\reference{}
Castro-Tirado, A. J. 1998, astro-ph/9803007
\reference{} Costa, E., et al. 1997, Nature, 387, 783
\reference{} Dwek, E. 1997, ApJ, submitted, astro-ph/9707024
\reference{} Eastman, R. G., Woosley, S. E., Weaver, T. A., \& Pinto, P. A. 
1994, ApJ, 430, 300
\reference{} 
Eichler, D., Livio, M., Piran, T., \& Schramm, D. N. 1989, Nature,
340, 126
\reference{}
Fenimore, E. E., Epstein, R. I., \& Ho, C. 1993, A\&AS, 97, 59
\reference{} 
Ferrini, F., Matteucci, F., Pardi, C., \& Penco, U. 1992, ApJ, 387, 138
\reference{}
Fishman, G. J., \& Meegan, C. A. 1995, ARA \& A, 33, 415
\reference{} 
Frail, D. A., et al. 1997, ApJ, 483, L91
\reference{} 
Fruchter, A. S., et al. 1998, 
to appear in Proc. of the 4th Huntsville GRB Symposium, 1997, Eds. C. A.
Meegan, R. Preece, \& T. Koshut,  astro-ph/9801169
\reference{} Galama, T. J., et al. 1997,  astro-ph/9712322
\reference{} Galama, T. J., et al. 1998, ApJL, in press, astro-ph/9802160
\reference{}
Garnavich, P. M., et al. 1998, ApJ, 493, L53
\reference{} 
Greggio, L., \& Renzini, A. 1983, A\&A, 118, 217
\reference{}
Haiman \& Loeb 1997a, ApJ, 483, 21 
\reference{}
---------------------. 1997b, ApJ, in press, astro-ph/9710208
\reference{}
Kirshner, R. 1998, private communication
\reference{} 
Klebasadel, R. W., Strong, I. B., \& Olson, R. A. 1973, ApJ, 182, L85
\reference{}
Loeb A. 1997, ``The First Stars and Quasars In the Universe'', in ``Science
With NGST'', Eds. E. Smith \& A. Koratkar, ASP Conf. Series, 133, 73;
astro-ph/9704290
\reference{} 
Madau, P. 1996, in ``Star Formation Near and Far'', 
AIP Conf. Proc. (New York: AIP), astro-ph/9612157
\reference{} 
Madau, P. 1997, to appear in ``The Hubble Deep Field'', 
ed. M. Livio, S. M. Fall, 
\& P. Madau, STScI Symposium Series, astro-ph/9709147
\reference{} 
Madau, P. 1998, to appear in ``The Young Universe: Galaxy Formation and 
Evolution at Intermediate and High Redshift'', 
ed. S. D'Odorico, A. Fontana, \& E. Giallongo, 
PASP, astro-ph/9801005
\reference{} 
Madau, P., Ferguson, H. C., Dickinson, M. E., Giavalisco, M., Steidel, C. C.,
\& Fruchter, A. 1996, MNRAS, 283, 1388
\reference{} 
Madau, P., Pozzetti, L., \& Dickinson, M. 1997, ApJ, submitted,
astro-ph/9708220
\reference{}
Mather, J., \& Stockman, P. 1996,
STScI newsletter, 13(2), 15 
\reference{} 
Meegan, C. A., et al. 1993, in Proc. of the Huntsville Gamma-Ray Burst 
Workshop, ed. G. Fishman, J. Brainerd, \& K. Hurley (New York: AIP), 3
\reference{} 
Metzger, M. R., et al. 1997, Nature, 387, 878
\reference{}
Miralda-Escud\'e, J., \& Rees, M. J. 1997, ApJ, 478, L57
\reference{} Mochkovitch, R., Hernanz, M., Isern, J., \& Martin, X. 1993, 
Nature, 361, 236
\reference{}
Narayan, R., Paczy\'nski, B., \& Piran, T. 1992, ApJ, 395, L83
\reference{} 
Nugent, P., Baron, E., Branch, D., Fisher, A., \& Hauschildt, P. H. 
1997, ApJ, 485, 812
\reference{} Paczy\'nski, B. 1986, ApJ, 308, L43
\reference{} Paczy\'nski, B. 1998, ApJ, 494, L45
\reference{} Pain, R., et al. 1996, ApJ, 473, 356
\reference{}
Perlmutter, S., et al. 1998, Nature, 391, 51
\reference{} Rhoads, J. E. 1997, ApJ, 487, L1
\reference{}
Rees, M. J. 1996, astro-ph/9608196
\reference{}
Rees, M. J. 1997, to appear in Proc. of the Texas symposium, Chicago, Dec.
1996, astro-ph/9701162
\reference{} Riess, A. G., Press, W. H., \& Kirshner, R. P. 1996, ApJ, 473, 88
\reference{} 
Ruiz-Lapuente, P., \& Canal, R. 1998, ApJL, 
in press, astro-ph/9801141
\reference{} 
Sadat, R., Blanchard, A., Guiderdoni, B., \& Silk, J. 1998, A\&A, 331, L96
\reference{} Salpeter, E. E. 1955, ApJ, 121, 161
\reference{}
Schaller, G., Schaerer, D., Meynet, G., \& Maeder, A. 1992, A\&AS, 96, 269
\reference{} Usov, V. V. 1992, Nature, 357, 472
\reference{} van Paradijs, J., et al. 1997, Nature, 386, 686
\reference{} Waxman, E. 1997, ApJ, 489, L33
\reference{} 
Wijers, R. A. M. J., Bloom, J. S., Bagla, J. S., \& Natarajan, P. 1997,
MNRAS, submitted, astro-ph/9708183
\reference{} Woods, E., \& Loeb, A. 1995, ApJ, 453, 583
\reference{} Woosley, S. E. 1993, ApJ, 405, 273
\reference{} 
Yungelson, L., \& Livio, M. 1998, ApJ, in press, astro-ph/9711201

\end{references}
\end{document}